%% file: ESQPTBorder.tex
\documentclass[10pt]{iopart}
\usepackage[utf8]{inputenc}
\usepackage{epsfig}
\usepackage{graphicx}
\usepackage{xcolor}
\usepackage{hyperref}
\usepackage{iopams}
\usepackage[toc,page]{appendix}
\usepackage{subcaption}
\newsavebox{\bigimage}

\usepackage{cite}

\hypersetup{
    colorlinks=true,
    linkcolor=blue,
    filecolor=magenta,      
    urlcolor=cyan,
    pdftitle={Excited-state quantum phase transitions in constrained systems},
    }

\graphicspath{{figures/}}

\def\commut#1#2{\left[{#1},{#2}\right]}
\def\abs#1{\left|{#1}\right|}
\def\norm#1{\left|\!\middle|{#1}\middle|\!\right|}
\def\ket#1{\left|{#1}\right\rangle}

\def\op#1{\hat{#1}}

\def\B{\op{B}}
\def\Bc{\op{B}^{\dagger}}
\def\b#1{\op{b}^{(#1)}}
\def\bc#1{\op{b}^{(#1)\dagger}}
\def\qo#1{\op{q}^{(#1)}}
\def\po#1{\op{p}^{(#1)}}
\def\q#1{q^{(#1)}}
\def\qs#1{q^{(#1)2}}
\def\p#1{p^{(#1)}}
\def\ps#1{p^{(#1)2}}
\def\vq#1{\vector{q}^{(#1)}}
\def\vp#1{\vector{p}^{(#1)}}
\def\vvq#1{\vector{\tilde{q}}^{(#1)}}
\def\vvp#1{\vector{\tilde{p}}^{(#1)}}
\def\s#1{\sigma^{(#1)}}
\def\ss#1{s^{(#1)2}}
\def\xv#1{\vector{x}^{(#1)}}

\def\M#1{\mathcal{M}^{(#1)}}
\def\H#1{\op{H}^{(#1)}}

\def\symp{\mathfrak{W}}
\def\coor#1{\eta_{#1}}
\def\ie{{\it i.e., }}
\def\cf{{\it cf. }}
\def\eg{{\it e.g., }}
\def\size{{\aleph}}

\def\bi#1{^{\scriptscriptstyle{({#1})}}}
\def\const{\mathrm{const.}}

\def\sign#1{\mathrm{sgn}\,{#1}\,}

\def\vector#1{\mathbf{#1}}
\def\vectorgreek#1{\boldsymbol{#1}}

\def\algebra#1{\mathfrak{#1}}
\def\group#1{\mathrm{#1}}

\graphicspath{{figures/}}

\begin{document}
\title{Excited-state quantum phase transitions in constrained systems}
\author{Jakub Novotný$^1$, Pavel Stránský$^1$ and Pavel Cejnar$^1$}
\address{$^1$ Institute of Particle and Nuclear Physics, Faculty of Mathematics and Physics, Charles University, V Hole\v{s}ovi\v{c}k\'ach 2, 18000 Prague, Czech Republic}

\ead{jakub.novotny@matfyz.cuni.cz}
\date{\today}

\begin{abstract}
We extend the standard semiclassical theory of Excited-State Quantum Phase Transitions (ESQPTs), based on a classification of stationary points in the classical Hamiltonian, to constrained systems.
We adopt the method of Lagrange multipliers to find all stationary points and their properties directly from the Hamiltonian constrained by an arbitrary number of integrals of motion, and demonstrate the procedure on an algebraic $\algebra{u}(3)$ boson model with two independent constraints.
We also elaborate the Holstein-Primakoff (HP) mapping, used to eliminate one degree of freedom in bosonic systems constrained by a conserved number of excitations, and address the fact that this mapping leads, in the classical limit, to a compact phase space with singular behaviour that conceals some stationary points at the phase space boundary.
It is shown that the HP method reveals all ESQPTs only after constructing a complete atlas of different HP mappings.

\end{abstract}

\submitto{\jpa}

\maketitle

\section{Introduction}
An Excited-State Quantum Phase Transition (ESQPT) is an extension of a Quantum Phase Transition (QPT), a phenomenon that manifests itself as nonanalytic properties of the ground state of a quantum system in the infinite-size limit when the system's external or internal coupling strengths are varied~\cite{Carr2011,Sachdev2011}, to the domain of excited energy states.
The ESQPTs are most pronounced in systems with a few (effective) degrees of freedom, which are often related to the collective dynamics of many-body systems.
They were first described in the interacting boson model of nuclear quadrupole vibrations~\cite{Cejnar2006a,Heinze2006,Caprio2008}, and later observed in a plethora of other collective systems, such as in the molecular vibron model~\cite{PerezBernal2010,Larese2011,Larese2013,KhaloufRivera2021}, in Bose-Einstein condensates~\cite{Shchesnovich2009,Relano2014}, in atom-field-interacting systems~\cite{PerezFernandez2011,Brandes2013,BastarracheaMagnani2014,Puebla2016,Kloc2017}, in periodically driven systems~\cite{Bastidas2014,Engelhardt2015}, and recently in exciton-polariton condensates and quantum superconducting circuits~\cite{PradoReynoso2023,ChavezCarlos2023,ChavezCarlos2024}.
For an exhaustive list of references and other details about the ESQPTs and their properties, we refer the reader to the recent review~\cite{Cejnar2021}.

Numerous studies targetting the ESQPTs from the semiclassical perspective have revealed a straightforward connection between the ESQPTs and the stationary points in the phase space of the system's classical Hamiltonian~\cite{Cejnar2008,BastarracheaMagnani2014,Stransky2014,Stransky2015b,Puebla2016,GarciaRamos2017}.
This connection has been fully established and a systematics of the ESQPTs has been provided in systems with nondegenerate stationary points~\cite{Stransky2016}. 
Degenerate stationary points can also be treated semiclassicaly, although they evade any general classification; nevertheless, a couple of theoretical case studies have been presented~\cite{Stransky2016,Kloc2017a}.

The first objective of this paper is to extend the semiclassical theory of ESQPTs to systems with constraints, which are most often induced by conserving quantities.
It employs the method of the Lagrange multipliers, which allows direct classification of the ESQPTs in constrained systems using the Lagrange function constructed from the Hamiltonian and all the constraints.
We show that this method is easy to use and robust, and its use circumvents the tedious search and application of a canonical transformation separating the constrained and unconstrained degrees of freedom.

A special class of models with constrained dynamics are fully-connected bosonic systems with an inherent constraint given by a conserving number of total boson excitations.
These systems have been successfully used to model collective dynamics in various many-body quantum systems, ranging from molecular physics~\cite{Iachello1981,Frank1994,Iachello1995,Iachello1996,Iachello2003}, nuclear physics~\cite{Iachello1987,Frank1994}, spin systems~\cite{Lipkin1965,Caprio2008,GarciaRamos2017}, and superconducting nonlinear oscillators~\cite{Iachello2023,Iachello2024}.
Traditional and routinely used techniques for their semiclassical analysis of ESQPTs involve the Holstein-Primakoff (HP) mapping~\cite{Holstein1940,Macek2019} or a method of an energy functional obtained via the coherent states~\cite{Gilmore1979,Dieperink1980,Zhang1990,PerezBernal2008}, both of which provide an explicit canonical transformation to eliminate the constrained degree of freedom.
Unfortunately, these mappings often lead to stationary points that appear at the boundary (or at infinity) of the resulting phase space~\cite{BastarracheaMagnani2014,Macek2019,Gamito2022}, which prevents the semiclassical classification of related ESQPTs.
This fact has even led to the conjecture that these ``boundary'' ESQPTs are of an entirely different nature~\cite{BastarracheaMagnani2014,Gamito2022}.
One particular attempt to treat the boundary exploits its hyperspherical shape by introducing local hyperspherical coordinates and proposes to analyse the boundary stationary points in the tangent subspace only~\cite{Macek2019}, but this procedure did not lead to a robust and practically applicable method.

The second objective of this paper is to demonstrate that the presence of the boundary stationary points in fully-connected bosonic systems is a consequence of singularity of the HP mapping and that a different HP transformation can map the same stationary point to the interior of the classical phase space, where the semiclassical ESQPT analysis can be performed.
Only a complete atlas of HP mappings, covering the full unconstrained phase space, is capable of revealing all ESQPTs in the bosonic systems.
However, the number of necessary HP mappings in the atlas grows with the number of bosonic degrees of freedom, and some of the mappings can even go against the internal algebraic structure of the models.
This makes the general Lagrange method more efficient in complicated bosonic systems.

The paper is organised as follows: 
Section~\ref{sec:ConstrainedSystems} provides a concise review of the semiclassical ESQPT analysis and introduces the method of Lagrange multipliers, which extends the ESQPT theory to general constrained systems.
Section~\ref{sec:BosonSystems} recalls the fully-connected bosonic models with conserving total number of boson excitations and shows a way how to construct a complete atlas of HP mappings in all available bosonic degrees of freedom.
It presents a detailed case study in the $\algebra{u}(3)$ boson model, where both Lagrange method and the HP mappings are used to reveal all ESQPTs and their properties.
The power of the Lagrange method is further demonstrated in a doubly constrained regime of the $\algebra{u}(3)$ model.
Finally, we conclude in Section~\ref{sec:Conclusions}.

\section{Semiclassical level density in general constrained systems}
\label{sec:ConstrainedSystems}
This section reviews the principal concepts of the ESQPT theory and summarises the connection between the ESQPTs and the stationary points of the corresponding classical Hamiltonian; more details on ESQPTs can be found in the recently published review~\cite{Cejnar2021}.
The ESQPT analysis is then extended to systems with an arbitrary number of constraints or integrals of motion by adopting the method of Lagrange multipliers. This extension constitutes the first original contribution of this paper.

\subsection{ESQPT singularities of level density}
\label{sec:ESQPTs}
We will consider a bound quantum system with $f$ degrees of freedom described on Hilbert space $\mathcal{H}$ by a time-independent Hamiltonian $\op{H}$ with a discrete energy spectrum $E_{n},n=1,2,\dots$, given by the solution of the stationary Schrödinger equation
$\op{H}\ket{\psi_{n}}=E_{n}\ket{\psi_{n}}$.
Precursors of the ESQPTs are primarily studied in the smooth level density
\begin{equation}\label{eq:smoothingdelta}
    \overline{\rho}(E)=\sum_{n}\overline{\delta}(E-E_{n}),
\end{equation}
where $\overline{\delta}$ is a $\delta$-like smoothing function satisfying 
\begin{equation}\label{eq:smoothingFunction}
    \int\overline{\delta}(y)dy=1,\quad
    \int y\overline{\delta}(y)dy=0,\quad
    \int y^2\overline{\delta}(y)dy=\Delta^{2}
\end{equation}
with the standard deviation $\Delta$ greater than the typical distance between neighbouring levels.

Semiclassically, the smooth level density arises from the energy derivative of the Liouville measure of the classical phase space and is given by the Weyl formula~\cite{Weyl1911,Gutzwiller1990}
\begin{equation}
    \label{eq:Weyl}
    \overline{\rho}(E)=\left(\frac{\size}{2\pi}\right)^{f}\frac{\partial}{\partial E}\int_{H(\vector{x})<E}d^{2f}\vector{x},
\end{equation}
where $H(\vector{x})$ is the classical counterpart of the Hamiltonian $\op{H}$, $\vector{x}=(\vector{q},\vector{p})$ specifies a point in the $2f$-dimensional phase space $\sigma$ of canonically conjugated positions $\vector{q}=(q_{1},\dots,q_{f})$ and momenta $\vector{p}=(p_{1},\dots,p_{f})$, and $\size$ is the size parameter of the system, often proportional to the reciprocal value of the Planck constant, $\size\propto 1/\hbar$.
Since the level density~\eref{eq:Weyl} diverges in the limit $\size\rightarrow\infty$, rescaling 
\begin{equation}
    \overline{\rho}_{\mathrm{cl}}(E)=\lim_{\size\rightarrow\infty}\frac{\overline{\rho}(E)}{\size^{f}}
\end{equation}
is necessary to obtain a finite semiclassical level density.

The smooth level density can develop ESQPT singularities in the classical limit even if the classical Hamiltonian $H(\vector{x})$ is a well-behaved function on $\sigma$.
The source of these singularities lies in the stationary points satisfying $\nabla H=0$.
If a stationary point $\vector{x}_{\mathrm{st}}$ is nondegenerate, {\it i.e.}, the Hessian matrix
\begin{equation}
    \label{eq:Hessian}
    D^2H_{kl}(\vector{x})\equiv\frac{\partial^{2}H(\vector{x})}{\partial x_{k}\partial x_{l}}
\end{equation}
has a nonvanishing determinant at $\vector{x} = \vector{x}_{\mathrm{st}}$, then the corresponding singularity can be uniquely classified by the number of degrees of freedom $f$ and the index of the nondegenerate stationary point $r$, which counts the number of negative eigenvalues of $D^2H$~\cite{Stransky2016}.
More specifically, the singularity will appear in the $(f-1)$-th derivative of $\overline{\rho}_{\mathrm{cl}}(E)$ in a form of a jump or a logarithmic divergence,
\begin{equation}
    \label{eq:ESQPTSingularities}
    \frac{\partial^{f-1}\overline{\rho}_{\mathrm{cl}}}{\partial E^{f-1}}\propto\left\{
        \begin{array}{ll}
            (-1)^{\frac{r}{2}}\Theta(E-E_{\mathrm{st}}), & r\ \mathrm{even},\\
            (-1)^{\frac{r+1}{2}}\ln\abs{E-E_{\mathrm{st}}}, & r\ \mathrm{odd},
        \end{array}
    \right.
\end{equation}
where $E_{\mathrm{st}}=H(\vector{x}_{\mathrm{st}})$ and $\Theta$ is the Heaviside step function.

Singularities induced by degenerate stationary points can be found similarly. 
However, apart from a few particular cases, their classification cannot be written in a compact form~\cite{Stransky2016}.

\subsection{General constrained dynamics}
\label{sec:Lagrange}
Let us suppose now that a quantum system is described by a Hamiltonian $\op{H}$ with $f+c$ degrees of freedom, where $c$ is the number of independent constraints $\{\op{\Phi}_{\alpha}\}_{\alpha=1}^{c}$ such that for all $\alpha$,
\begin{equation}\label{eq:quantumconstraints}
    \op{\Phi}_{\alpha}|\psi_{c}\rangle = 0,\quad \forall |\psi_{c}\rangle \in \mathcal{H}_{c}\subset\mathcal{H} ,
\end{equation}
where $\mathcal{H}_{c}$ denotes a subspace of physical states of the constrained system;
the quantum constraints act on the full Hilbert space $\mathcal{H}$ and annihilate the states from~$\mathcal{H}_{c}$.
The~constraint often originates in an integral of motion $\op{I}_\alpha$ satisfying $\commut{\op{H}}{\op{I}_\alpha}=0$. 
In that case 
\begin{equation}
    \op{\Phi}_{\alpha}(I_{\alpha})\equiv\op{I}_\alpha-I_{\alpha},
\end{equation}
where $I_{\alpha}$ is a concrete eigenvalue of operator $\op{I}_\alpha$, upon which the constraint explicitly depends.
Note that this is a stronger condition than the conservation of $\op{I}_\alpha$ in $\mathcal{H}$, since it does not allow for superpositions of states with different eigenvalues $I_\alpha$.

In the classical limit, the resulting classical Hamiltonian $H(\vector{X})$ and the constraints $\Phi_\alpha(\vector{X})$---classical counterparts of operators $\op{\Phi}_{\alpha}$---are functions of canonical coordinates $\vector{X} = (\vector{Q},\vector{P})$ on a $2(f+c)$-dimensional phase space $\Omega$.
We will consider only well-behaved constraints satisfying $\nabla\Phi_\alpha(\vector{X})\neq 0$ with linearly independent gradients in the vicinity of the constrained surface 
\begin{equation}
    \Phi_\alpha(\vector{X})=0.    
\end{equation}
The classical constraints restrict the dynamics to a $(2f + c)$-dimensional subspace $\Sigma\subset\Omega$, further referred to as the \emph{constrained} phase space.
There are additional $c$ cyclic coordinates $\varphi_{\alpha}$ in $\Sigma$ conjugated to the constraints that can be determined via the Hamilton-Jacobi theory and separated by an appropriate canonical transformation.
Therefore, relevant dynamics can be studied in a $2f$-dimensional \emph{reduced} subspace $\sigma\subset\Sigma$.
If the constraints are given by integrals of motion, then the canonical transformation leads to the action-angle coordinates in the corresponding degrees of freedom with trivial dynamics~\cite{Gutzwiller1990}.

\begin{figure}[h]\centering
    \includegraphics[width=0.75\linewidth]{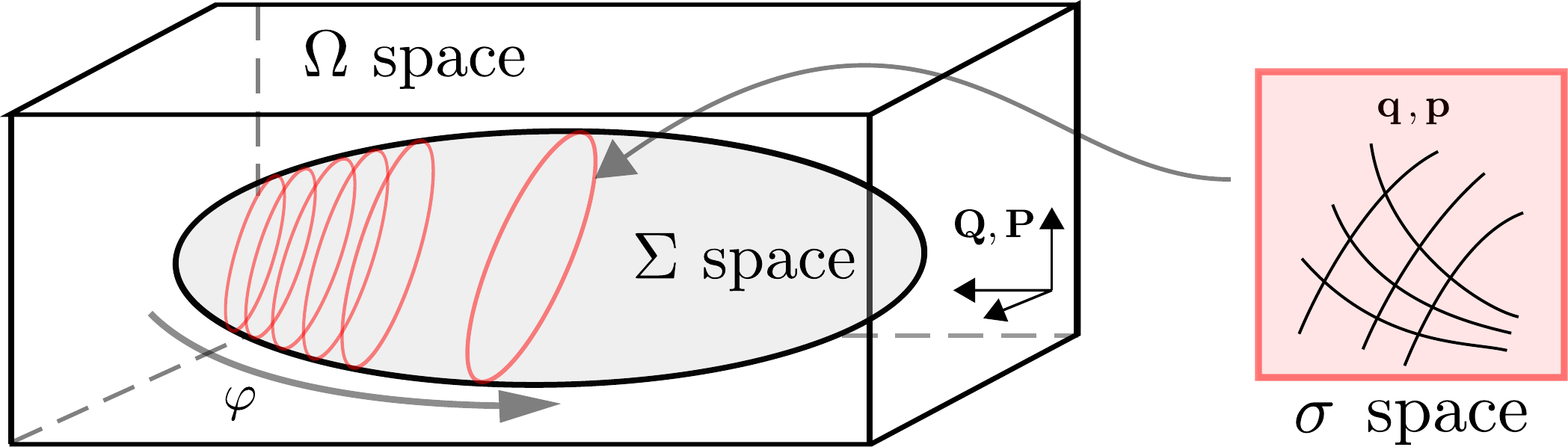}
    \caption{
        Relation between the full (unconstrained) phase space $\Omega$, the constrained space $\Sigma$, and the reduced space $\sigma$.
        The dynamics of the physical system with $c$ constraints, occurring in the $2(f+c)$-dimensional space $\Omega$ with coordinates $(\vector{Q},\vector{P})$, is confined to the $(2f+c)$-dimensional space~$\Sigma$.
        The relevant degrees of freedom are described in the $2f$-dimensional space $\sigma$ with coordinates $(\vector{q},\vector{p})$; this space covers the constrained space $\Sigma$ under the trivial evolution of the cyclic coordinates $\{\varphi_{\alpha}\}_{\alpha=1}^{c}$ conjugated to the constraints.
    }
    \label{fig:drawing}
\end{figure}

The resulting classical Hamiltonian $H^{(\sigma)}(\vector{x})$, where $\vector{x}=(\vector{q},\vector{p})$ are coordinates on~$\sigma$, describes the same physical system as $H(\vector{X})$ stripped of all degrees of freedom related to the given constraints.
Schematic relations between the subspaces are displayed in Figure~\ref{fig:drawing}.

We assume that the system is initially described by a Hamiltonian with constraints.
If, instead, the system were given by a singular Lagrangian, \ie a Lagrangian whose second-derivative matrix with respect to the velocities is singular, the Dirac-Bergmann algorithm~\cite{Dirac1964,Rothe2010} must be applied to derive the corresponding Hamiltonian and identify the associated constraints.
Note that performing the classical limit of a constrained system represents the inverse of Dirac's approach to the quantisation of constrained systems, a method widely used in gauge theories. In this context, the values of the cyclic coordinates are fixed by a choice of the gauge~\cite{Henneaux1992}.

\subsection{Constrained level density via the Lagrange method}
Finding and classifying the ESQPTs in the constrained system can be done by explicit elimination of the constrained degrees of freedom and a subsequent analysis of the stationary points of the classical Hamiltonian $H^{(\sigma)}$ in space $\sigma$.
This can be, in general, a challenging procedure, see, \eg~\cite{Kloc2017a}, where an explicit canonical transformation in a concrete system is provided.
However, stationary points of $H^{(\sigma)}(\vector{x})$ with their proper indexes $r$ can be found directly from $H(\vector{X})$ in the full space $\Omega$ using the Lagrange function
\begin{equation}
    \label{eq:LagrangeFunction}
    L(\vector{X},\vectorgreek{\lambda}) = H(\vector{X}) + \sum_{\alpha=1}^{c}\lambda_{\alpha}\Phi_{\alpha}(\vector{X}),
\end{equation}
where $\vectorgreek{\lambda}=(\lambda_{1},\dots,\lambda_{\alpha})$ are Lagrange multipliers corresponding to the constraints $\vector{\Phi}=(\Phi_{1}\dots,\Phi_{\alpha})$~\cite{Bertsekas1996}.
A stationary point $\vector{X}_{\mathrm{st}}$ of $H$ in $\Omega$ under given constraints is a solution of
\begin{equation}\label{eq:stationaryPointL}
    \nabla_{(\vector{X},\vectorgreek{\lambda})} L(\vector{X},\vectorgreek{\lambda}) = 0,
\end{equation}
where the gradient $\nabla_{(\vector{X},\vectorgreek{\lambda})}$ is taken with respect to the coordinates $(\vector{X},\vectorgreek{\lambda})$, or equivalently by solving
\begin{equation}
    \nabla_{\vector{X}}H(\vector{X})=0,\quad\vector{\Phi}(\vector{X})=0.
\end{equation}

As proven in Appendix~\ref{app:proof}, $\vector{X}_{\mathrm{st}}$ has its counterpart $\vector{x}_{\mathrm{st}}$---a stationary point of $H^{(\sigma)}$ in $\sigma$---and gives the same energy,
\begin{equation}\label{eq:firstStatement}
    E_{\mathrm{st}}\equiv H^{(\sigma)}(\vector{x}_{\mathrm{st}})=H(\vector{X}_{\mathrm{st}})=L(\vector{X}_{\mathrm{st}},\vectorgreek{\lambda}_{\mathrm{st}}).
\end{equation}
The index of the stationary point $r(\vector{x}_{\mathrm{st}})$ given by the negative eigenvalues of the Hessian $D^{2}H^{(\sigma)}$ at $\vector{x}_{\mathrm{st}}$ satisfies
\begin{equation}\label{eq:secondStatement}
    r(\vector{x}_{\mathrm{st}}) = r_\Sigma(\vector{X}_{\mathrm{st}}),        
\end{equation}
where $r_\Sigma(\vector{X}_{\mathrm{st}})$ is determined from the Hessian of the Lagrange function $L$ at $\vector{X}_{\mathrm{st}}$ restricted to the $\Sigma$ subspace, $D^2L|_{\Sigma}$, in which the derivatives are taken only in directions tangent to $\Sigma$, \ie the directions perpendicular to the constraints given by vectors $\nabla\Phi_{\alpha}$ must be excluded.
Note that the exact transformation from coordinates $\vector{X}$ to $\vector{x}$ is not necessary for the determination of the energy $E_{\mathrm{st}}$ and index $r$, and the method of the Lagrange multipliers does not provide a prescription how to find this transformation.
The technical details on the evaluation of the restricted Hessian and a discussion of the degenerate stationary point cases are provided in Appendix~\ref{app:proof}.

It remains to show that the smooth level density of the constrained system in~$\sigma$ corresponds to the smooth level density $\overline{\rho}_{\Omega}(E)$ on $\Omega$ with constraints, calculated directly from Hamiltonian $H(\vector{X})$.
Including the constraints into~\eref{eq:Weyl} leads to
\begin{equation}
    \overline{\rho}_{\Omega}(E)=\left(\frac{\size}{2\pi}\right)^{f+c} \frac{\partial}{\partial E} \int_{H(\vector{X}) < E} d^{2 (f + c)}\vector{X}\prod^{c}_{\alpha = 1}\delta(\Phi_{\alpha}(\vector{X})),
\end{equation}
where $\delta(x)$ is the Dirac delta function.
If we perform a partial integration in the special coordinate system obtained from the Hamiltonian coordinates by a canonical transformation separating the constraints $\Phi_{\alpha}$ and conjugated cyclic coordinates $\varphi_{\alpha}$, see~\eref{eq:xphiPhi}, and use the integral over conjugate coordinates in one degree of freedom $\int d\varphi_{\alpha}\,d\Phi_{\alpha}\,\delta(\Phi_{\alpha})=2\pi/\aleph$~\cite{Gutzwiller1990}, we get
\begin{eqnarray}
\overline{\rho}_{\Omega}(E)&=\left(\frac{\size}{2\pi}\right)^{f+c} \frac{\partial}{\partial E} \int_{H'(\vector{x},\vector{\Phi}) < E} {d}^{2f}\vector{x}\,{d}^{c}\vector{\varphi}\,{d}^{c}\vector{\Phi}\prod^{c}_{\alpha = 1}\delta(\Phi_{\alpha})\nonumber\\
&=\left(\frac{\size}{2\pi}\right)^{f}\frac{\partial}{\partial E} \int_{H'(\vector{x},\vector{\Phi}=\vector{0}) < E} {d}^{2f} \vector{x}  \nonumber\\
&=\overline{\rho}(E),
\end{eqnarray}
which is exactly the smooth level density~\eref{eq:Weyl} for $H^{(\sigma)}$.

To summarise this section, the ESQPTs in a constrained system can be found by analysing the stationary points $(\vector{X}_{\mathrm{st}},\vectorgreek{\lambda}_{\mathrm{st}})$ of the Lagrange function $L$; the energies of the ESQPTs are given by~\eref{eq:firstStatement} and the nature of the ESQPT nonanalyticity can be discerned from the index $r_\Sigma$ of the restricted Hessian via Eq.~\eref{eq:ESQPTSingularities}.
The singularity induced by a nondegenerate stationary point appears in the $(f-1)$-th derivative of the semiclassical level density.

\section{ESQPTs in fully-connected bosonic systems}
\label{sec:BosonSystems}
    This section introduces algebraically formulated bosonic systems with a natural constraint originating in the conservation of the total number of boson excitations, and focuses on the HP mapping---an explicit prescription to eliminate the constrained degree of freedom.
    We recall the construction of the HP mapping and deal in a completely new way with the difficulties arising from the singular behaviour of the classical dynamics at the boundary of the resulting reduced phase space.
    We discuss that these obstacles will be overcome by a complete atlas of various HP mappings that cover the full unconstrained phase space.
    The analysis given in this section represents a major extension and generalisation of the approach described in~\cite{Macek2019},
    and constitutes the second original contribution of this paper.

\subsection{Fixed-$N$ bosonic systems}
\label{sec:BosonModels}
    Let us consider a quantum system described by a fully-connected bosonic Hamiltonian with one- and two-body interactions,
    \begin{equation}
        \label{eq:Hamiltonian_Second_Quantisation}
            \op{H}=\sum_{k,l=0}^f \varepsilon_{kl} \Bc_{k}\B_{l}
            +\sum_{k,k',l,l'=0}^f\nu_{k k' l l'} \Bc_{k}\Bc_{k'}\B_{l}\B_{l'},
    \end{equation}
    where $\{\Bc_{k},\B_{k}\}_{k=0}^{f}$ stand for $f+1$ pairs of creation and annihilation operators satisfying the standard boson commutation relations and $\varepsilon_{kl},\nu_{kk'll'}$ are appropriately chosen coefficients in order to obtain a Hermitian Hamiltonian that scale properly with the system size.
    The Hamiltonian can be expressed solely by operators $\op{\mathfrak{g}}_{kl}=\Bc_{k}\B_{l}$, which are generators of the $\group{U}(f+1)$ group and form the $\algebra{u}(f+1)$ dynamical algebra of the system~\cite{Iachello2015}.
    The linear Casimir operator of the dynamical algebra
    \begin{equation}
        \label{eq:N}
        \op{N}=\sum_{k=0}^{f}\underbrace{\Bc_{k}\B_{k}}_{N_{k}},
    \end{equation}
    corresponds to the total number of boson excitations and commutes with the Hamiltonian, $\commut{\op{H}}{\op{N}}=0$, hence acts as an integral of motion.
    Its eigenvalue $N$ labels the totally symmetric irreducible representations of the dynamical algebra.
    The physical state is always described by a definite number of boson excitations~$N$, so that it belongs to a single irreducible representation labelled by $N$ (no physical state of the system can be a superposition of states with different total numbers of excitations).
    Conserving $N$ generates the constraint 
    \begin{equation}
        \label{eq:OPhiN}
        \op{\Phi}_{N}\equiv\hat{N}-N.
    \end{equation}    
    
    The Hamiltonian~\eref{eq:Hamiltonian_Second_Quantisation} can be cast in the position-momentum form by introducing $2(f+1)$ operators $\{\op{Q}_k,\op{P}_k\}_{k=0}^{f}$ of canonically conjugate positions and momenta via the transformation~\cite{Macek2019}
    \begin{equation}
        \label{eq:BosonicOpPQ}
        \Bc_{k}=\sqrt{\frac{N}{2}}\left(\op{Q}_k-i \op{P}_k\right), \qquad 
        \B_{k}=\sqrt{\frac{N}{2}}\left(\op{Q}_k+i \op{P}_k\right).
    \end{equation}   
    Operators $\op{Q}_{k},\op{P}_{l}$ satisfy commutation relations $\commut{\op{Q}_{k}}{\op{P}_{l}}=\frac{i}{N} \delta_{kl}$ ($\delta_{kl}$ is the Kronecker delta), from which the reciprocal value of the number of excitations can be identified as an effective Planck constant, hence $N$ serves as the size parameter, $\size\equiv N$.
    The infinite-size limit 
    \begin{equation}
        \label{eq:ClassicalLimit}
        \lim_{N\rightarrow\infty}\frac{\op{H}}{N}\mapsto H(\vector{Q},\vector{P})
    \end{equation}
    provides the classical limit of the system, in which $\vector{Q}=(Q_{0},Q_{1},\dots,Q_{f}),\vector{P}=(P_{0},P_{1},\dots,P_{f})$ become continuous commuting variables, spanning a classical $2(f+1)$-dimensional phase space $\Omega$~\cite{Yaffe1982}.
    The classical limit of the constraint~\eref{eq:OPhiN} is given by
    \begin{equation}
        \label{eq:PhiN}
        \lim_{N\rightarrow\infty}\frac{\op{\Phi}_{N}}{N}\mapsto\Phi_{N}(\vector{Q},\vector{P})=\frac{1}{2}\underbrace{\sum_{k=0}^{f}\left(Q_{k}^2 + P_{k}^2\right)}_{R_{\Sigma}^{2}}-1=0,
    \end{equation}
    so that the classical dynamics is confined to a $(2f+1)$-dimensional sphere $\Sigma$ with radius $R_{\Sigma}=\sqrt{2}$ embedded in $\Omega$.
    Note that the quantum energies must be given in units of $N$, \ie per one boson excitation, for a straightforward comparison between quantum and classical calculations.

    We will demonstrate the HP mapping and the procedure of finding the ESQPTs in a concrete $f=2$ boson model built on the dynamical algebra $\algebra{u}(3)$ with Hamiltonian 
    \begin{equation}
        \label{eq:U(3)hamiltonian}
            \op{H}=\left(1 - \xi\right)\op{C}_{1}[\algebra{u}(2)]-\frac{\xi}{N+1}\op{C}_{2}[\algebra{o}(3)]-\epsilon\op{D},
    \end{equation}
    where
    \begin{eqnarray}
        \fl \op{C}_{1}[\algebra{u}(2)]&=&\Bc_1\B_1 + \Bc_2\B_2,\nonumber\\
        \fl \op{C}_{2}[\algebra{o}(3)]&=&\left( \Bc_1\B_0 - \Bc_0\B_1 \right)^2 + \left( \Bc_2\B_0 - \Bc_0\B_2 \right)^2 +\left( \Bc_2\B_1 - \Bc_1\B_2 \right)^2
    \end{eqnarray}
    are the linear and quadratic Casimir operators of two subalgebras $\algebra{u}(2)$ and $\algebra{o}(3)$, respectively, of the dynamical algebra $\algebra{u}(3)$. 
    The normalization factor in front of the $\op{C}_{2}[\algebra{o}(3)]$ term guarantees a perfect agreement between the range of the quantum spectrum and the available energies in the classical limit because the eigenvalues of $\op{C}_{2}[\algebra{o}(3)]$ range from $0$ to $N+1$~\cite{Iachello1996}.
    The tunable parameter $\xi\in[0,1]$ describes a transition between the $\algebra{u}(2)$ and $\algebra{o}(3)$ symmetries with a QPT at $\xi=1/5$~\cite{PerezBernal2008}.
    The dipole operator
    \begin{equation}
        \op{D}= \Bc_2\B_0 - \Bc_0\B_2
    \end{equation}
    models the interaction with an external (electromagnetic) field.
    Its strength is governed by the second tunable parameter $\epsilon$ of the Hamiltonian.
    If the dipole operator is not present, \ie $\epsilon=0$, the system has $\group{O}(2)$ symmetry, and the quadratic Casimir operator 
    \begin{equation}
        \label{eq:l2}
        \op{l}^{2}\equiv\op{C}_{2}[\algebra{o}(2)]=\left( \Bc_2\B_1 - \Bc_1\B_2 \right)^2
    \end{equation}    
    of the $\algebra{o}(2)$ algebra serves as the third integral of motion besides the energy and $N$, making the system integrable.
    For general nonzero values of $\epsilon$ and $\xi$, the system is chaotic~\cite{Novotny2023}.

    The transformation to positions and momenta~\eref{eq:BosonicOpPQ} and the limit $N\rightarrow\infty$ lead to the classical Hamiltonian on the full three-dimensional space $\Omega$,
    \begin{eqnarray}
        \label{eq:U(3)classicalHam}
        \fl H(\vector{Q},\vector{P}) = &\frac{1-\xi}{2}\left(Q_1^2 + P_1^2 + Q_2^2 + P_2^2 \right)\nonumber\\
        \fl&-\xi\left[\left(P_1Q_0-P_0Q_1 \right)^2 + \left(P_1Q_2-P_2Q_1 \right)^2 + \left(P_2Q_0-P_0Q_2 \right)^2  \right]\nonumber\\
        \fl&-\epsilon\left(P_2Q_0-P_0Q_2\right);
    \end{eqnarray}
    it is a polynomial of the fourth order in the phase-space coordinates $Q_{k},P_{k}$.

    Note that this $\algebra{u}(3)$ model describes bending vibrational modes of linear polyatomic molecules~\cite{Iachello1996,Larese2011,Larese2013,KhaloufRivera2021,KhaloufRivera2022a}, often referred to as the vibron model in this context, but also the Bose-Einstein condensates composed of spin 1 constituents~\cite{Kunkel2018,Kunkel2019,Rautenberg2020,Niu2023}.
    It has also proven to serve as a valuable tool for probing and validating theoretical concepts like quantum critical phenomena and quantum chaos~\cite{PerezBernal2008,Larese2011,KhaloufRivera2022,KhaloufRivera2022a,Novotny2023}.

\subsection{Holstein-Primakoff mappings}
\label{sec:HPmapping}
    A particular way to construct a classical limit of the boson Hamiltonian~\eref{eq:Hamiltonian_Second_Quantisation} is the HP mapping, introduced and elaborated in~\cite{Holstein1940,Blaizot1978,Macek2019}. 
    A concise review and some new technical extensions of the HP mapping are given in Appendix~\ref{app:HPmapping}.
    Here we recall just the essential features of the HP mapping and discuss its limitations in the context of the constrained systems.

    At the quantum level, the HP mapping uses the conserving number of total boson excitations $N$ to eliminate one specific boson type (often labelled by $j=0$ and called the scalar boson~\cite{Cejnar2007}).
    We generalise the procedure and consider HP mappings in any boson type $j$.
    It consists in the transformation of boson operators $\{\Bc_{k},\B_{k}\}_{k=0}^{f}\mapsto\{\bc{j}_{k},\b{j}_{k}\}_{k=0}^{f}$, after which the pair $\bc{j}_{j},\b{j}_{j}$ only encodes the conserving number $N$ and hence can be eliminated from the resulting Hamiltonian $\H{j}$~\cite{Macek2019}, reducing effectively the number of degrees of freedom from $f+1$ to $f$.
    The elements of the $\algebra{u}(f+1)$ algebra map as follows:
    \begin{eqnarray}
        \label{eq:productOfTwoBBoperators}
        \Bc_{k}\B_{l}&=&\bc{j}_{k}\b{j}_{l}\quad\textrm{for $k,l\neq j$},\nonumber\\        
        \Bc_{k}\B_{j}&=&\bc{j}_{k}\sqrt{N-\sum_{l\neq j}\bc{j}_{l}\b{j}_{l}}\quad\textrm{for $k\neq j$},\nonumber\\
        \Bc_{j}\B_{j}&=&N-\sum_{l\neq j}\bc{j}_{l}\b{j}_{l}.
    \end{eqnarray}            

    The classical limit is performed by following the same steps as for $\Bc_{k},\B_{k}$: the boson operators $\{\bc{j}_{k},\b{j}_{k}\}_{j\neq k}$ are substituted by canonically conjugated position and momentum operators $\{\qo{j}_{k},\po{j}_{k}\}_{j\neq k}$ following~\eref{eq:BosonicOpPQ}, after which the limit $N\rightarrow\infty$ leads to a classical Hamiltonian $H^{(j)}$ on the reduced $2f$-dimensional phase space $\s{j}$ spanned by coordinates
    \begin{eqnarray}
        \vvq{j}&=\left(\q{j}_{0},\dots,\q{j}_{j-1},\q{j}_{j+1},\dots,\q{j}_{f}\right),\\
        \vvp{j}&=\left(\p{j}_{0},\dots,\p{j}_{j-1},\p{j}_{j+1},\dots,\p{j}_{f}\right).
    \end{eqnarray}
    Explicit expressions for the classical HP mapping are provided in Appendix~\ref{app:HPmapping}.

    In coordinates $(\vvq{j},\vvp{j})$, the constraint~\eref{eq:PhiN} translates to
    \begin{eqnarray}
        \label{eq:Ballx}
        \ss{j}=\sum_{k\neq j}\left[\qs{j}_{k} + \ps{j}_{k}\right]
        \leq2,
    \end{eqnarray}
    so that the reduced space $\s{j}$ is a compact $2f$-dimensional ball with a boundary at radius $R_{\sigma}=\sqrt{2}$.
    As shown in Appendix~\ref{app:HPmapping}, the classical HP mapping preserves the relative number of excitations of each boson type,
    \begin{equation}
        \label{eq:nk}
        n_{k}=\frac{N_{k}}{N}=\frac{Q_{k}^2+P_{k}^2}{2}=\frac{\left(\q{j}_{k}\right)^2+\left(\p{j}_{k}\right)^2}{2},j\neq k,   
    \end{equation}
    where $N_{k}$ is an eigenvalue of the number operator $\op{N}_{k}=\Bc_{k}\B_{k}=\bc{j}_{k}\b{j}_{k}$.
    This fact helps discuss and visualize the dynamics in spaces $\Sigma$ and $\s{j}$: for each pair of position and momentum, one can introduce a single coordinate
    \begin{equation} 
        \label{eq:EtaCoordinates-particle_space}
        \coor{k}\equiv\sign{\q{j}_{k}}\sign{\p{j}_{k}}\sqrt{2n_{k}}=\sign{Q_{k}}\sign{P_{k}}\sqrt{2n_{k}},
    \end{equation} 
    where $\sqrt{2n_{k}}$ equals the norm of both vectors $\vector{X}_{k}\equiv(Q_{k},P_{k})$ and $\xv{j}_{k}\equiv(\vq{j}_{k},\vp{j}_{k})$, and $\sign{x}$ is the signum function equal to $-1$ for $x<0$ and $+1$ for $x\geq 0$.
    The classical HP mapping $\M{j}:\Sigma\mapsto\s{j}$ can therefore be interpreted as a projection from the sphere $\Sigma$ to the ball $\s{j}$ along the direction given by $\eta_{j}$ coordinate axis.
    
    A so far under-discussed weakness of the mapping $\M{j}$ is its discontinuity in the phase space region corresponding to states with a vanishing number of $j$-th boson excitation, \ie the region with $N_j = 0$, whose image forms a boundary of $\s{j}$.
    The origin of the discontinuity is presented in Appendix~\ref{app:HPmapping} and demonstrated in Figure~\ref{fig:pathFIN} by an example of a simple continuous path (not necessarily a trajectory) in the $f=2$ boson system. 
    The path, marked by the oriented green line, is chosen to cross the subspace with no $j=0$ boson excitations in $\Sigma$, \ie the subspace with $N_{0}=0$.
    It is parametrised by varying $Q_0$ while keeping the number of the $j=2$ boson excitations constant, $N_2 = \const$:
    \begin{eqnarray}\label{eq:path}
        &Q_0 \in [1,-1],Q_1 = \sqrt{\frac{3}{2} - Q_0^2},Q_2 = \frac{1}{\sqrt{2}},\nonumber\\
        &P_0=P_1=P_2 = 0.
    \end{eqnarray}
    This path satisfies the constraint~\eref{eq:PhiN}, hence lies in the restricted space $\Sigma$.

    The path starts at a specific initial point in $\Sigma$, namely at $(Q_{1},Q_{2},P_{1},P_{2})=(\sqrt{3/2},\sqrt{1/2},0,0)$, which maps to $(\q{0}_{1},\q{0}_{2},\p{0}_{1},\p{0}_{2})=(\sqrt{3/2},\sqrt{1/2},0,0)$ in~$\s{0}$.
    It then continues smoothly in $\Sigma$ and exhibits no discontinuity or other singularity; see Figure~\ref{fig:pathFIN}(a).
    However, as the path approaches $Q_{0}=0$, it reaches the boundary of the reduced space $\s{0}$ where the mapping $\M{j}$ is discontinuous due to $\norm{X_{0}}=0$, see Eq.~\eref{eq:qkpk}.
    At the boundary, the signs of both $\q{0}_{1}$ and $\q{0}_{2}$ flip, hence the path appears in the opposite quadrant in the $(\eta_{1},\eta_{2})$ plane, as shown in Figure~\ref{fig:pathFIN}(b).
    The image of this path is continuous in both $\s{1}$ and $\s{2}$ (not shown in the Figure) as it does not cross regions without $j=1$ and $j=2$ boson types, respectively.    

    \begin{figure}[!hbtp]
        \centering
            \includegraphics[width=\linewidth]{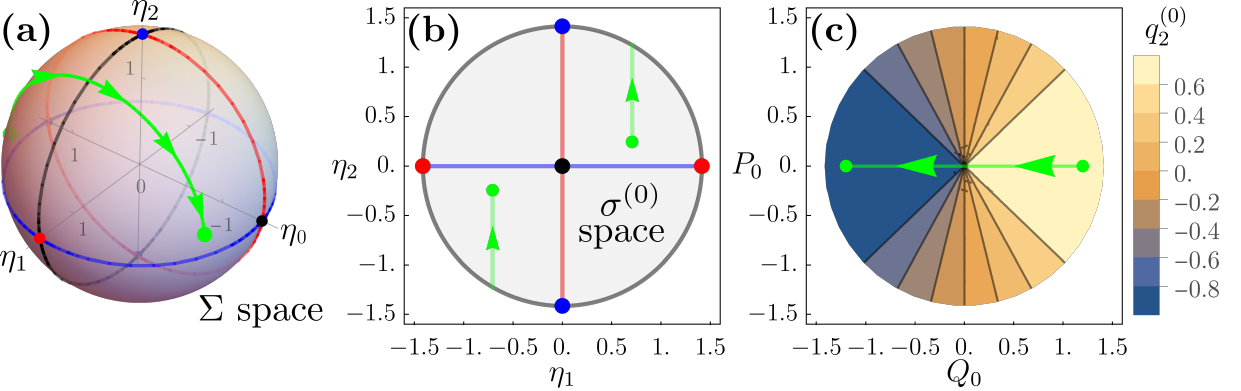}
            \caption{(a) A schematic example of a continuous path (green solid oriented line) given by Eq.~\eref{eq:path} in space $\Sigma$ of a general three-boson system.
            (b) The path mapped onto space $\s{0}$ by the $\M{0}$ mapping.
            Both $\Sigma$ and $\s{0}$ are visualized in coordinates~$\eta_{k}$.
            Coloured bullets mark states populated by just one type of boson (black: $j=0$ boson, red: $j=1$ boson, blue: $j=2$ boson).
            Lines of corresponding colours indicate states \emph{not} populated by the given boson type.
            The discontinuity of the path in $\s{0}$ is a consequence of the sign change of $\q{0}_{1}$ and $\q{0}_{2}$.
            (c) The contour graph shows $\q{0}_{2}$ as a function of $Q_0$ and $P_0$ for constant $Q_2 = \sqrt{1/2},P_2 = 0$, see Eq.~\eref{eq:qkpk}.
            The central point $Q_0=P_0=0$ is mapped to the whole boundary of $\s{0}$.
            }
            \label{fig:pathFIN}
    \end{figure}

    The discontinuity of the HP mapping prevents one from finding the stationary points at the boundary via the vanishing gradient of Hamiltonian $\H{j}$ and analysing their properties using the spectrum of the Hessian matrix since both $\nabla{H}$ and $D^{2}H$ are not defined there.
    At the level on the HP mappings, this problem can be resolved either by constructing the whole atlas of $\M{j},j=0,\dots,f$, or by resewing the discontinuous open neighbourhood on the boundary of an individual $\s{j}$, which is in fact equivalent to finding a canonical transformation to the HP mapping in another boson type, in other words choosing another map from the atlas.
    As a result, each stationary point will appear inside at least one of the spaces $\s{j}$, allowing for a correct identification and classification of the ESQPT singularities.
    
    Note that another way of performing the classical limit of the boson system, restricted by conserving $N$, employs various definitions of coherent states~\cite{Gilmore1979,Dieperink1980,Zhang1990,PerezBernal2008}.
    It often leads to classical Hamiltonians parametrised by coordinates in which the boundary of the reduced phase space corresponding to $N_{j}=0$ is projected to infinity.
    This means that the affected stationary points lie also at infinity, which again prevents their proper analysis.
    
    \subsection{ESQPTs via the HP mappings}
    The atlas that allows us to find and classify all the stationary points of the $\algebra{u}(3)$ model Hamiltonian~\eref{eq:U(3)hamiltonian} consists of three classical HP mappings onto reduced phase spaces $\sigma^{(j)},j=0,1,2$, each obtained by removing one boson type.
    The resulting classical Hamiltonians are
    \begin{eqnarray}
       \label{eq:H(0)}
       \fl H^{(0)} = &\frac{1-\xi}{2}s^{(0)2} \nonumber\\
       \fl&-\xi \left[\left(\ps{0}_1 + \ps{0}_2\right)^2 \left(2 - s^{(0)2}\right)^2 + \left(\p{0}_2 \q{0}_1-\p{0}_1 \q{0}_2 \right)^2  \right]\nonumber\\
       \fl&-\epsilon \p{0}_2\sqrt{2 - s^{(0)2}},\\
       \label{eq:H(1)}
       \fl H^{(1)} = &\frac{1-\xi}{2}\left(2 - \qs{1}_0 - \ps{1}_0\right)\nonumber\\
       \fl & -\xi \left[\left(\ps{1}_0 + \ps{1}_2\right)^2 \left(2 - s^{(1)2}\right)^2 + \left(\p{1}_2 \q{1}_0-\p{1}_0 \q{1}_2 \right)^2  \right]\nonumber\\
       \fl&-\epsilon \left(\p{1}_2 \q{1}_0-\p{1}_0 \q{1}_2 \right),\\
       \label{eq:H(2)}
       \fl H^{(2)} = &\frac{1-\xi}{2}\left(2 - \qs{2}_0 - \ps{2}_0\right)\nonumber\\
       \fl&-\xi \left[\left(\ps{2}_0 + \ps{2}_1\right)^2 \left(2 - s^{(2)2}\right)^2 + \left(\p{2}_1 \q{2}_0-\p{2}_0 \q{2}_1 \right)^2  \right]\nonumber\\
       \fl&-\epsilon \p{2}_0\sqrt{2 - s^{(2)2}},
    \end{eqnarray}
    where $s^{(j)2}$ are the radial coordinates in the phase spaces $\s{j}$~\eref{eq:Ballx}.
            
    The stationary points $\vector{x}_{\mathrm{st}}$ are solutions of
    \begin{equation}\label{eq:GradHamOnsigma}
        \nabla H^{(j)}(\vector{x}) = 0
    \end{equation}
    in the inner regions $\ss{j} < 2$.
    Their indices $r$ are given by the number of negative eigenvalues of the corresponding Hessian matrices $D^2H^{(j)}$ \eref{eq:Hessian}; if the Hessian matrix had at least one zero eigenvalue, the stationary point would be degenerate.

    \begin{figure}[!htb]
        \centering
            \includegraphics[width=0.8\linewidth]{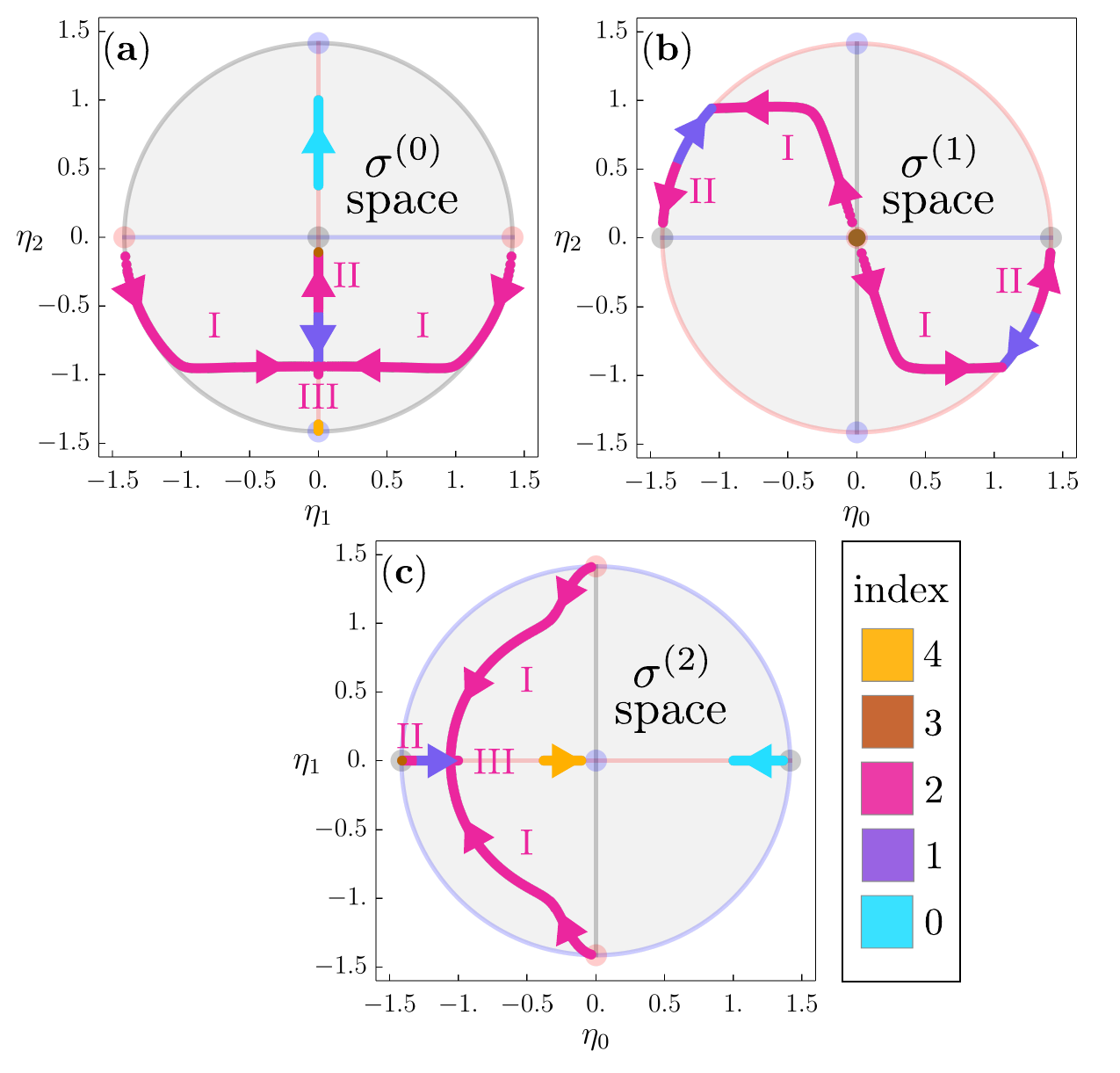}
            \caption{
            Stationary points of the $\algebra{u}(3)$ Hamiltonians $H^{(j)}$ in their corresponding phase spaces $\s{j},j=0,1,2$, parametrised by the coordinates $\coor{k}$~\eref{eq:EtaCoordinates-particle_space}.
            Model parameters are $\epsilon = 0.3$ and $\xi \in [0,1]$.
            Varying $\xi$ creates paths of stationary points in the phase space; 
            their directions for increasing $\xi$ are marked by arrows.
            Line colours indicate the indexes $r$ of the stationary points, see the legend on the right.
            Roman numerals enumerate the segments of the $r=2$ stationary points shown in Figure~\ref{fig:duo}; segment~$\mathrm{III}$ appears as a short line following the $r=1$ (violet) path.
            }
            \label{fig:sigmaExample}
    \end{figure}

    Figure~\ref{fig:sigmaExample}(a)---(c) shows the positions of stationary points given by the solution of~\eref{eq:GradHamOnsigma} for each of the Hamiltonians~\eref{eq:H(0)}---\eref{eq:H(2)}, respectively.
    The strength of the external field is fixed at $\epsilon=0.3$ and the remaining parameter is varied in the range $\xi \in [0,1]$.
    As the parameter $\xi$ increases, the stationary points move in the phase space along oriented lines in the directions indicated by arrows.
    
    First, let us focus on the stationary point corresponding to the ground state.
    In the classical limit, the ground state sits at the global minimum of the Hamiltonian, which is a nondegenerate stationary point with index $r=0$ (cyan line in Figure~\ref{fig:sigmaExample}).
    In the quantum calculation, the ground state is populated only by $j=0$ and $j=2$ boson excitations.
    Therefore, this stationary point can be revealed, and its index determined, only in spaces $\s{0}$ and $\s{2}$; in $\s{1}$ it would be mapped onto the diverging boundary.
    
    In contrast, there is a stationary point populated solely by $j=1$ bosons excitations, \ie appearing only in $\s{1}$ and detectable only from $\H{1}$. 
    This stationary point is at position $\q{1}_{0}=\p{1}_{0}=\q{1}_{2}=\p{1}_{2}=0$ independent of the parameter $\xi$, resulting in an ESQPT at $E=1-\xi$, see~\eref{eq:H(1)}.
    Its index is $r=3$ for $0.023\leq\xi\leq 0.977$ (brown point in the centre of Figure~\ref{fig:sigmaExample}(b)), and $r=2$ elsewhere.
    Limiting to the analysis of $\s{0}$ only, as is usually done~\cite{BastarracheaMagnani2014,Macek2019}, and omitting the HP mapping onto $\s{1}$ would result in a failed attempt to identify this ESQPT in the spectrum of the quantum system with some stationary point of the classical Hamiltonian. 

    The rest of the stationary points (violet, magenta and yellow lines in Figure~\ref{fig:sigmaExample}) are populated by all three types of bosons and can be found in all three spaces $\s{j}$.
    There are three different stationary points with the same index $r=1$, which can be discerned by the Roman numerals.
    Note that some of the stationary points seemingly appear to lie at the boundary, for example the violet points in Figure~\ref{fig:sigmaExample}(b); they are in fact located close to the boundary, but still in the inner region of the reduced phase space.
    This will be clearly apparent later in Figure~\ref{fig:ball}.
          
    \begin{figure}
        \centering
        
        \sbox{\bigimage}{%
          \begin{subfigure}[b]{.45\textwidth}
          \centering
          \includegraphics[width=\textwidth]{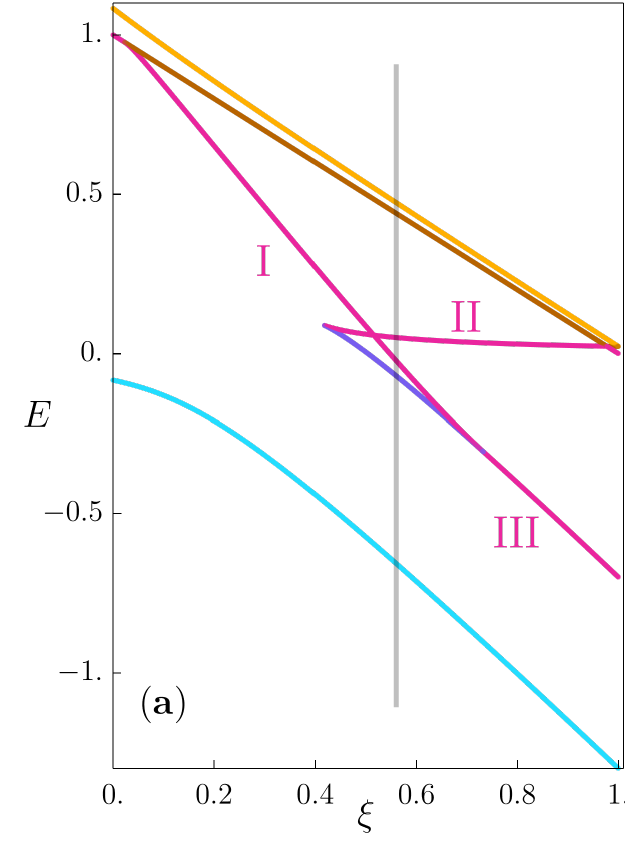}
        
          \vspace{0pt}
          \end{subfigure}%
        }
        
        \usebox{\bigimage}\hfill
        \begin{minipage}[b][\ht\bigimage][s]{.5\textwidth}
          \begin{subfigure}{\textwidth}
          \centering
          \includegraphics[width=\textwidth]{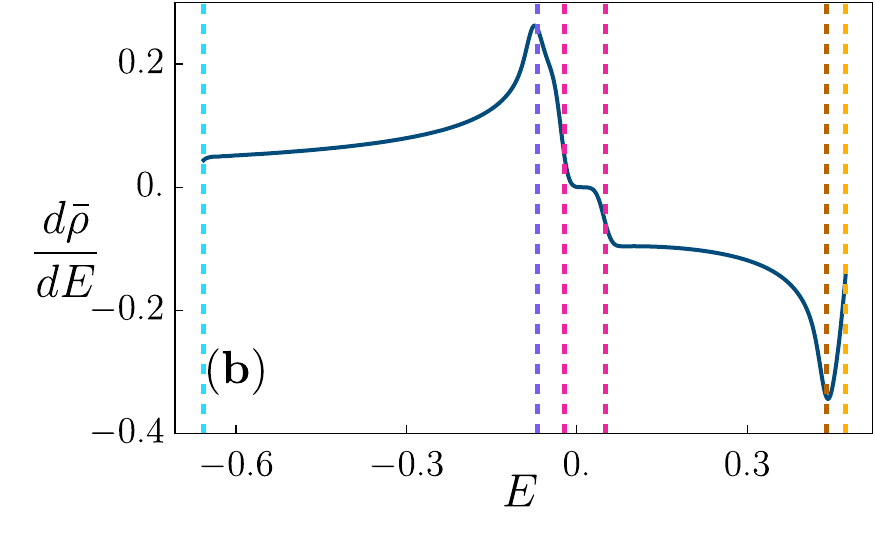}
          \end{subfigure}%
          \vfill
          \begin{subfigure}{\textwidth}
          \centering
          \includegraphics[width=\textwidth]{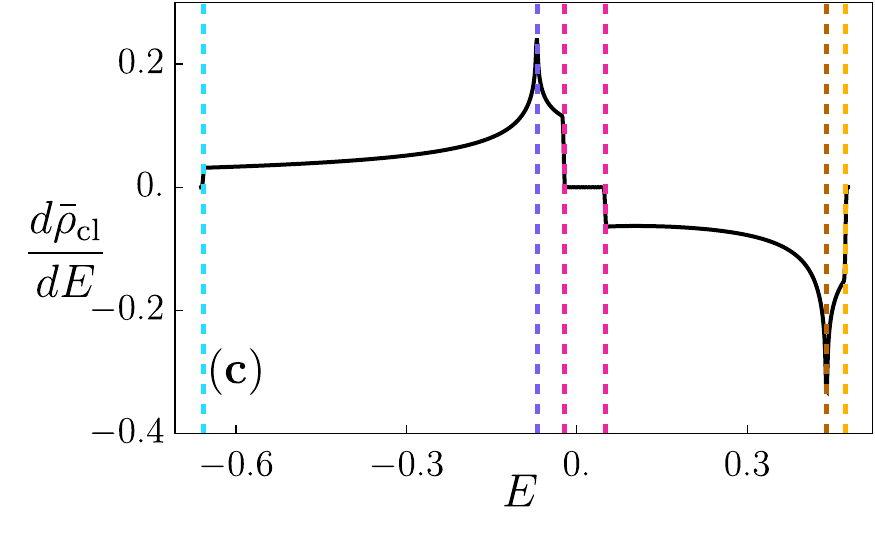}
          \end{subfigure}
        
          \vspace{0pt}
        \end{minipage}
        
        \caption{
            (a) Energies of all the stationary points of the $\algebra{u}(3)$ Hamiltonians $H^{(j)}$ obtained from the complete atlas of HP mappings. Model parameters are $\epsilon = 0.3$ and $\xi \in [0,1]$.
            The colours of the lines encode the indexes of the stationary points according to the colour scale given in Figure~\ref{fig:sigmaExample}.
            (b) First derivative of the smooth level density obtained by smoothing the quantum spectrum for $\xi=0.56$ (the value is indicated by the grey vertical line in panel (a)) by a Gaussian smoothing function $\overline{\delta}(y)=1/\sqrt{2\pi\Delta^{2}}\exp\left(-y^2/2\Delta^2\right)$ with variance $\Delta^2 = 0.0075$.
            The size parameter is $N = 150$. The quantum energies are in units of $N$.
            (c) First derivative of the semiclassical level density calculated via the Weyl formula~\eref{eq:Weyl}.
            }
            \label{fig:duo}
        \end{figure}

    Energies $E_{\mathrm{st}}=H^{(j)}(\vector{x}_{\mathrm{st}})$ of all the stationary points calculated in the reduced spaces $\s{j}$ are shown in Figure~\ref{fig:duo}(a) as functions of parameter $\xi$; the colour code is the same as in Figure~\ref{fig:sigmaExample}.
    We observe a rich ESQPT structure: besides the global minimum ($r=0$) and maximum ($r=4$) of the Hamiltonian, there are also intertwined lines of stationary points with indexes $r=1,2,3$ (the indexes were determined from the Hessian matrices $D^2H^{(j)}$).
    The Roman numerals indicate corresponding segments of the $r=2$ stationary point in Figure~\ref{fig:sigmaExample}.

    The constrained quantum system has $f=2$ degrees of freedom.
    Since all the stationary points are nondegenerate, the corresponding ESQPT singularities appear in the first energy derivative of the smooth level density $\overline{\rho}$ according to~\eref{eq:ESQPTSingularities}, as shown in Figures~\ref{fig:duo}(b) and~\ref{fig:duo}(c).
    While panel (b) displays $d\overline{\rho}/dE$ calculated directly from the Gaussian-smoothed quantum spectrum obtained by a numerical diagonalization of the Hamiltonian matrix~\eref{eq:U(3)hamiltonian} with parameters $\xi = 0.56, \epsilon = 0.3$ and $N = 150$, panel (c) contains $d\overline{\rho}_{\mathrm{cl}}/dE$ for $\overline{\rho}_{\mathrm{cl}}$ calculated semiclassically using the Weyl formula~\eref{eq:Weyl} for Hamiltonian $\H{0}$.
    Dashed vertical lines in both panels (b) and (c) mark the energies of the stationary points displayed in panel (a).
    Due to the finite size of the quantum system, panel (b) displays smooth precursors of the ESQPTs rather than sharp singularities, which are obtained only in the infinite-$N$ limit.
    
    The observed singularities are entirely in agreement with~\eref{eq:ESQPTSingularities}:
    upward jumps connected with stationary points with index $r=0$ and $r=4$ (global minimum and maximum of the Hamiltonian, respectively), logarithmic divergences associated with $r=1$ (pointing upwards, second singularity from the left) and $r=3$ (pointing downwards, second singularity from the right), and downward jumps corresponding to $r=2$ (two singularities in the middle of the graph).

    \subsection{ESQPT analysis via the Lagrange multipliers}
    The Lagrange function \eref{eq:LagrangeFunction} for Hamiltonian $H(\vector{Q},\vector{P})$ of the $\algebra{u}(3)$ model~\eref{eq:U(3)classicalHam} constrained by the number of total boson excitations~\eref{eq:PhiN} reads as
    \begin{equation}
        \label{eq:LagrangeFunctionExample}
        L_{1}(\vector{Q},\vector{P},\lambda_{N}) = H(\vector{Q},\vector{P}) + \lambda_{N}\left[\frac{1}{2}\sum_{k = 0}^2 (Q_k^2 + P_k^2) - 1 \right].
    \end{equation}
    Stationary points of $L_{1}$ are solutions of 7 coupled equations~\eref{eq:stationaryPointL} for positions and momenta $\vector{Q},\vector{P}$ and Lagrange multiplier $\lambda_{N}$.
    Due to the action-angle structure of the constrained coordinates, the stationary points are degenerated in the cyclic variable and each of them form a one-dimensional region spanned by a free parameter, see Appendix~\ref{app:proof}.
    However, the ESQPT energies of the stationary points does not depend on the choice of the free parameter, and its value can thus be set arbitrarily.

    \begin{figure}[h]
        \centering
            \includegraphics[width=0.8\linewidth]{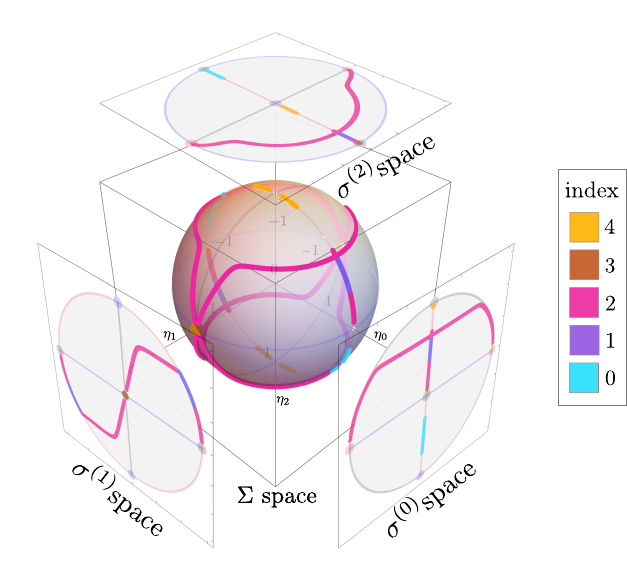}
            \caption{Stationary points of Lagrange function~\eref{eq:LagrangeFunctionExample} in $\coor{k}$ coordinates, in which the phase space $\Sigma$ has the form of a $2$-sphere.
            Model parameters are the same as in Figure~\ref{fig:sigmaExample}.
            The stationary points form lines parametrised by $\xi$.
            The indexes of the stationary points, obtained from the restricted Hessian $D^2L_{1}|_{\Sigma}$, are indicated by colours (see the legend).
            For comparison, there are also plotted the three restricted phase spaces $\s{j}$ with stationary points of $H^{(j)}$ from Figure~\ref{fig:sigmaExample}.
            We see that the classical HP mapping $\Sigma\mapsto\s{j}$ acts as a projection.
            }
            \label{fig:ball}
    \end{figure}

    Stationary points for the $\algebra{u}(3)$ model with $\epsilon=0.3$ and $\xi\in[0,1]$ are plotted in Figure \ref{fig:ball} in coordinates $\coor{k}$ defined by~\eref{eq:EtaCoordinates-particle_space}.
    In these coordinates, the restricted space $\Sigma$ forms a 2-dimensional sphere.
    The indexes of the stationary points calculated from the restricted Hessian $D^{2}L_{1}|_{\Sigma}$ are marked in the same colours as in Figure~\ref{fig:sigmaExample}.
    The energies $E=H(\vector{X}_{\mathrm{st}})$ of the stationary points give the same graph as shown in Figure~\ref{fig:duo}(a).
    Both approaches thus lead to equivalent results.
    
    Besides the stationary points determined by the Lagrange multipliers method on the sphere $\Sigma$, we also provide the HP ``projections'' to all the three available reduced spaces $\s{j}$.
    It is evident that some lines of stationary points present in $\Sigma$ are absent in the $\s{j}$ spaces.
    This comes from the ambiguity connected with the signum function in the definition~\eref{eq:EtaCoordinates-particle_space} of coordinates $\eta_{k}$, which causes multiple lines in $\Sigma$ to map to a single line in $\s{j}$, and has no observable consequences.

    The Lagrange method offers several advantages for the ESQPT analysis compared to the explicit elimination of the dependent degrees of freedom via the HP mappings.
    Firstly, it evades the need for the complicated HP coordinate transformations~\eref{eq:qkpk}, which can be tedious to construct, especially in algebraic systems based on higher-rank dynamical algebras and $j\neq0$ bosons.
    Secondly, unless we analyse the whole atlas of HP mappings, we cannot be sure that we have found all the stationary points.
    Thirdly, the HP mapping usually contaminates the reduced Hamiltonians with square roots, see~\eref{eq:H(0)} and~\eref{eq:H(1)}, and the stationary points can only be found numerically.
    In contrast, the Lagrange function for an algebraically formulated Hamiltonian in the form~\eref{eq:Hamiltonian_Second_Quantisation} is a polynomial in coordinates $(\vector{Q},\vector{P},\lambda_{N})$, which enables the solution of \eref{eq:stationaryPointL} to be found in an analytical form (not given here), even when the model parameters are not specified.
    The existence of the full analytical solution also provides an upper limit to the number of ESQPTs in the system.
    Fourthly, the Lagrange method allows for a more profound analysis of the properties of the Hamiltonian manifold.
    One can actually solve equations~\eref{eq:stationaryPointL} in the complex domain.
    Then, at some values of the external parameters, the solutions move from the complex numbers to the real numbers or vice versa, resulting in a sudden change in the number of stationary points.
    This usually happens at the ``vertices'' or branching of the lines of stationary points, see \eg the two lines (one with $r=1$ and other with $r=2$) in Figure~\ref{fig:duo}(a) that emerge at $\xi=\xi_e\approx 0.42$:
    for $\xi<\xi_{e}$ there are two solutions of~\eref{eq:stationaryPointL} with a nonzero imaginary part, while for $\xi>\xi_{e}$ they get to the real plane and become true stationary points.

    \subsection{ESQPTs in multiply constraint systems}
    \label{sec:MultipleConstraints}
    The algebraic $\algebra{u}(3)$ model allows us to demonstrate the method of Lagrange multipliers in case of multiple constraints because, for some values of the tunable parameters $(\xi,\epsilon)$, the model has an additional integral of motion besides conserving~$N$.
    We focus on the configuration with $\epsilon=0$, for which the $\algebra{o}(2)$ Casimir invariant~\eref{eq:l2} is conserved, $[\op{H},\op{l}^2]=0$.
    We only consider eigenstates $\op{l}^2$ with a given eigenvalue $l^2$, which means that the relevant Hilbert space is an irreducible subspace $\mathcal{H}_{l}\subset\mathcal{H}_{N}$.
    The corresponding constraint is 
    \begin{equation}
        \op{\Phi}_{l} = \op{l}^{2} - l^{2}.
    \end{equation}
    Note that in the context of bending modes of linear molecules, $\op{l}^{2}$ corresponds to the 2D angular momentum, whereas in the Bose-Einstein condensate, it specifies the total magnetization of the condensate.

    The classical limit of the additional constraint in space $\Omega$ reads as 
    \begin{equation}\label{eq:classLimitl2}
        \lim_{N\rightarrow\infty}\frac{1}{N}\op{\Phi}_{l}\mapsto
        \Phi_{l}=l^{2}(\vector{Q},\vector{P})-\ell^{2}
        =(P_1Q_2 - P_2Q_1)^2-\ell^{2},
    \end{equation}
    where $\ell^{2}\in[0,1]$, and the stationary points are determined from the Lagrange function 
    \begin{eqnarray}
        \label{eq:LagrangeFunctionExamplel}
        L_{2}(\vector{Q},\vector{P},\lambda_{N},\lambda_{l}) = H(\vector{Q},\vector{P})&+ \lambda_N \left[\frac{1}{2}\sum_{k = 0}^2 (Q_k^2 + P_k^2) - 1\right]\nonumber\\
        & + \lambda_{l} \left[(P_1Q_2 - P_2Q_1)^2 -\ell^2 \right],
    \end{eqnarray}
    by solving~\eref{eq:stationaryPointL} for $\vector{Q},\vector{P}, \lambda_N$ and $\lambda_{l}$.

    \begin{figure}
        \begin{subfigure}[h]{0.5\textwidth}
            \includegraphics[width=\textwidth]{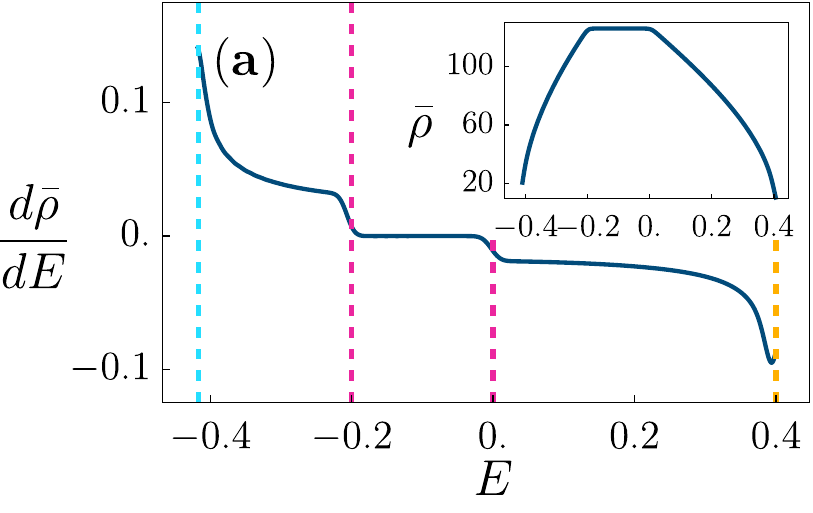}
        \end{subfigure}
        \begin{subfigure}[h]{0.48\textwidth}
            \includegraphics[width=\textwidth]{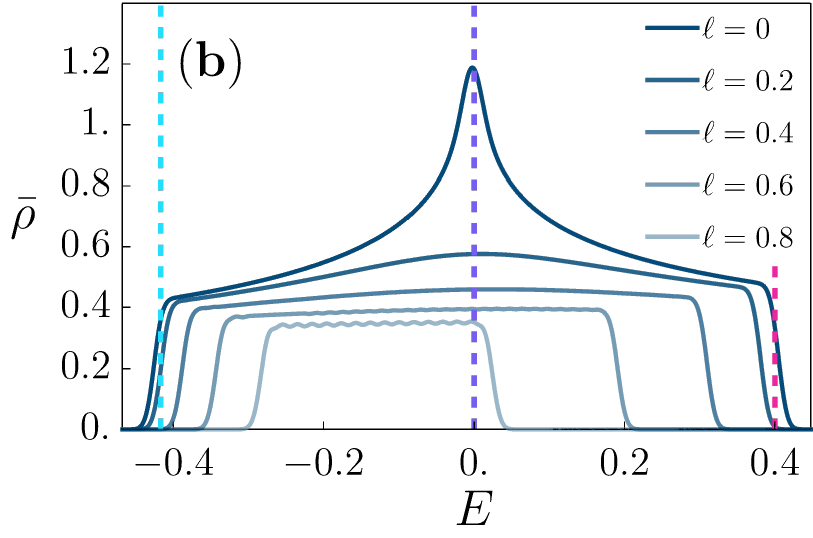}
        \end{subfigure}
        \caption{
            Smooth level density $\overline{\rho}(E)$ obtained from a numerical diagonalization of the $\algebra{u}(3)$ Hamiltonian with $\xi = 0.6$, $\epsilon = 0$. (a) First derivative ${d}\overline{\rho}/{d}E$ ($\overline{\rho}(E)$ is plotted in the inset) when the constraint $\Phi_{l}$ is switched off.
            (b) Constraint $\Phi_{l}$ switched on and $\ell = \{0,0.2,0.4,0.6,0.8\}$ subspaces of eigenstates shown.
            The positions and indexes of the stationary points were calculated using the Lagrange multipliers.
            They are shown by dashed vertical lines and the indexes are encoded in the same colours as in Figure~\ref{fig:ball}.
            The size parameter and the Gaussian-smoothing parameter are $N=150$ and $\Delta = 0.01$, respectively.
            All energies are in the units of $N$.
        }
        \label{fig:ESQPT_l}
    \end{figure}

    A numerical example of the integrable $\algebra{u}(3)$ model with parameters $(\xi,\epsilon)=(0.6,0)$ is given in Figure~\ref{fig:ESQPT_l}.
    It shows the Gaussian-smoothed level densities $\overline{\rho}$ of the quantum energy levels obtained by a diagonalization of the Hamiltonian~\eref{eq:U(3)hamiltonian} with $N=150$.
    Panel (a) corresponds to the system with active $\Phi_{N}$ while constraint $\Phi_{l}$ is switched off.
    Besides the global minimum and maximum, there are two stationary points inducing ESQPT singularities, manifested at $E = -0.2$ and $E = 0$ as downward jumps (index $r=2$) in the first derivative of the smooth level density due to effective $f=2$ degrees of freedom.    
    Panel (b) shows $\overline{\rho}(E)$ for the system further constrained by $\Phi_{l}$ for various values of $\ell$.
    The doubly constrained system has effectively only $f=1$ degree of freedom, so that the ESQPTs are expected to be in the smooth level density itself.
    An ESQPT singularity appears only in the $\ell = 0$ subspace at energy $E = 0$ (upward logarithmic divergence corresponding to $r=1$), as already revealed in~\cite{Santos2015}, while the other subspaces with higher values of $\ell$ show no ESQPT singularities.
    The positions of the stationary points were calculated using the Lagrange multipliers and are marked with horizontal 
    dashed lines; their indexes were obtained from $D^2L_{1}|_{\Sigma}$ (panel a) and $D^2L_{2}|_{\Sigma}$ (panel b).
    
    If we compare both panels of Figure~\ref{fig:ESQPT_l}, we see that the additional constraint $\Phi_{\ell}$ changes both the number and the character of the ESQPTs compared to the case with a single constraint $\Phi_{N}$.
    The ESQPT missing in the doubly constrained case is connected with an invariant subspace with maximal value $\ell_{\mathrm{max}}= 1$ (corresponding to eigenvalue $l_{\mathrm{max}}=N$).
    This subspace is formed by only one state with energy $E_{l_{\mathrm{max}}}=1-2\xi=-0.2$ (hence not shown in the Figure), $\dim\mathcal{H}_{l_{\mathrm{max}}} = 1$, and limits from above the lowest energies of all the available subspaces $\mathcal{H}_{l}$.
    Therefore, the ESQPT at $E=-0.2$ can be considered as a result of an interplay of all the subspaces limited by $\mathcal{H}_{l_{\mathrm{max}}}$.

    There are no other stationary points of $L_{2}$ for $0<\ell^2<1$ except for global minima and maxima. 
    Note that if we extend $L_{2}$ into the complex domain, the stationary point of $L_{2}$ that generates the ESQPT in the $\ell^2 = 0$ subspace move to complex values of $\vector{X}$ coordinates for all $\ell^2 > 0$ subspaces.
    These complex stationary points do not induce any ESQPT singularity; however, there presence near the real axis is responsible for the nonmonotonous behaviour of the smooth level density around $E\approx0$.

\section{Conclusions}
\label{sec:Conclusions}
We have presented a straightforward and easy-to-implement semiclassical method of finding and classifying the ESQPTs in the spectrum of a quantum system with any number of constraints induced by conserving quantities, based on the Lagrange multipliers, and demonstrated it on a case study of the algebraic $\algebra{u}(3)$ model.
Using the Lagrange method, we have found a complete set of stationary points and explained in full detail the rich ESQPT structure.
We have discussed a case with multiple conserved quantities (multiple constraints), corresponding to an integrable version of the model.
It has been demonstrated that additional constraints can nontrivially change the number and the character of the ESQPTs:
Firstly, new constraints effectively reduce the number of degrees of freedom, hence move the ESQPT singularities to lower derivatives of the smooth energy level density.
Secondly, the indexes of the ESQPTs may change, as they are determined from the Hessian restricted to the subspace of all constraints.
And finally, some ESQPTs may even be absent in the more constrained system, either because the associated stationary points lie only in subspaces with some particular values of the conserved quantities, or because they are a result of an interplay of all the subspaces. 

We have also examined the HP mappings in the algebraic bosonic systems with a conserving number of total boson excitations.
It has been shown that a single HP mapping (or an analogous transformation to eliminate the integral of motion) is not sufficient for a complete semiclassical analysis of the ESQPTs in the spectrum of a constrained system, since some of the stationary points can be hindered by the singular boundary of the reduced phase space.
At the classical level, the HP mapping can be considered as a projection of the constrained $(2f+1)$-dimensional sphere-shaped subspace of the full $2(f+1)$-dimensional phase space onto a reduced $2f$ dimensional ball-like phase space.
The presence of stationary points at the boundary is a consequence of this projection, and the corresponding ESQPTs are in no way different from other ESQPTs.
For a complete ESQPT analysis, it is necessary to construct an atlas of HP mappings to cover the whole unconstrained phase space and look for stationary points in each of the local coordinate maps.
This can be a demanding process, especially in systems with more degrees of freedom and with a more complicated algebraic structure.
We have performed the HP mappings individually for all available bosons in the $\algebra{u}(3)$ model, resulting in three different classical Hamiltonians living in three different projections of the full phase space, and obtained the same results as with the Lagrange method.

If we were to compare both methods, the former can be applied to any constrained system, while the latter is suitable only for a rather special class of bosonic systems with not too much complicated algebraic structure.
An advantage of the HP mapping is that it provides an explicit form of the canonical transformation separating the constrained degree of freedom.
Lagrange method does not offer such a canonical transformation, as it is not required for the ESQPT analysis at all.
The Lagrange function is directly constructed from the Hamiltonian and constraints, which are often polynomials in generalised coordinates, and therefore solving the equations for stationary points is computationally less demanding and can often even be found analytically. 
The simplicity of finding the stationary points is, to some extent, compensated by more involving (although straightforward) procedure of determining their indexes from the restricted Hessian matrix.
Therefore, the Lagrange method is more versatile, more robust and can significantly simplify the ESQPT analysis in complicated constrained systems in the future.

\section*{Acknowledgements}
The authors appreciate fruitful discussions with F.~Pérez-Bernal, J.~Střeleček and L.~Honsa.
The work was supported by The Charles University Grant Agency (grant no. 215323).
    
\begin{appendices}
    \section{Evaluation of the restricted Hessian}\label{app:proof}
    In this appendix we address statements~\eref{eq:firstStatement} and~\eref{eq:secondStatement} and discuss the construction of the restricted Hessian of the Lagrange function $L$.

    As stated in Section~\ref{sec:Lagrange}, the constraints (often related to integrals of motion) of a Hamiltonian system with $f+c$ degrees of freedom are conserved by the Hamiltonian flow on the full phase space $\Omega$.
    They are associated with $c$ conjugated pairs of local action-angle-like variables $(\Phi_{\alpha},\varphi_{\alpha})$.
    It is convenient to construct a local coordinate system at each point of $\Omega$ in the following way,
    \begin{equation}
        \label{eq:xphiPhi}
        \vector{X}' = (\vector{x},\varphi_1,\dots, \varphi_c,\Phi_1,\dots, \Phi_c),
    \end{equation}
    where $\vector{x}=(x_1,\dots,x_f)$ correspond to coordinates on the reduced phase space $\sigma$ of the constrained dynamics.
    The transformation $\mathcal{C}:\vector{X}\mapsto\vector{X}'$ is a canonical transformation, leading to Hamiltonian $H'(\vector{X}')$ in the new coordinates.
    The stationary points $\vector{X}'_{\mathrm{st}}$ of $H'$ on the restricted subspace $\Sigma$ with fixed $\Phi_{\alpha}$ are given by the solutions of equations
    \begin{equation}
    \label{eq:H'sigma}
    \nabla H'|_{\Sigma} = \left(\frac{\partial H'}{\partial x_1},\dots,\frac{\partial H'}{\partial x_f},\frac{\partial H'}{\partial \varphi_1}\dots,\frac{\partial H'}{\partial \varphi_c}\right)=\vector{0}.
    \end{equation}
    Since $H'$ does not depend on the cyclic variables, $H'=H'(\vector{x},\Phi_{1},\dots,\Phi_{c})$, then $\partial H'/\partial\varphi_\alpha = 0$ for all $\alpha=1,\dots,c$.
    Therefore,
    \begin{equation}
        \nabla_{\vector{x}}H'\equiv \left(\frac{\partial H'}{\partial x_1},\dots,\frac{\partial H'}{\partial x_f}\right)=\vector{0}
    \end{equation}
    and $\vector{X}'_{\mathrm{st}}$ determines a stationary point of $H^{(\sigma)}$,
    \begin{eqnarray}
        H^{(\sigma)}(\vector{x}_\mathrm{st})
        &\equiv H'(\vector{x}_\mathrm{st},\Phi_{\alpha} = 0)
        =H'(\vector{X}'_{\mathrm{st}})\nonumber\\
        &=H(\mathcal{C}^{-1}(\vector{X}'_{\mathrm{st}}))
        =H(\vector{X}_{\mathrm{st}}),
    \end{eqnarray}
    which is the statement~\eref{eq:firstStatement}.
    The sufficient condition for the stationary point~\eref{eq:H'sigma} can be expressed using the Lagrange function~\eref{eq:stationaryPointL}.
    
    Let us construct now the restricted Hessian necessary for evaluating the index of the stationary point~\eref{eq:secondStatement}.
    The Hessian matrix of $L'=H'+\sum\lambda_{\alpha}\Phi_{\alpha}$ in coordinates~\eref{eq:xphiPhi} can be expanded as
    \begin{equation}\label{eq:hessianL}
        D^{2}L'=\left(\begin{array}{ccc} 
            \left(D^{2}_{\vector{x}}L'\right)_{2f\times2f}
            &(0)_{2f\times c} 
            &\left(\frac{\partial^{2}L'}{\partial x_{k}\partial\Phi_{\alpha}}\right)_{2f\times c} \\
             (0)_{c\times2f} & (0)_{c\times c} & (0)_{c\times c}\\
             \left(\frac{\partial^{2}L'}{\partial\Phi_{\alpha}\partial x_{l}}\right)_{c\times 2f} &(0)_{c\times c} & \left(\frac{\partial^{2}L'}{\partial\Phi_{\alpha}\partial\Phi_{\beta}}\right)_{c\times c}\end{array}\right),
    \end{equation}
    where the symbol $(\bullet)_{a\times b}$ denotes a submatrix of size $a\times b$; the zero submatrices $(0)_{a\times b}$ come from the independence of $H$ on the cyclic variables.
    The index of the stationary point is determined from the canonical coordinates~\eref{eq:xphiPhi}, so that the Lagrange multipliers obtained from~\eref{eq:stationaryPointL} enter as fixed parameters.
    The Hessian on the reduced space 
    \begin{eqnarray}
        D^{2}_{\vector{x}}L'&\equiv\left(\frac{\partial^{2}L'}{\partial x_{k}\partial x_{l}}\right)_{2f\times2f}\nonumber\\
        &=\Bigg(\frac{\partial^{2}H'}{\partial x_{k}\partial x_{l}}+\sum_{\alpha}\lambda_{\alpha}\underbrace{\frac{\partial^{2}\Phi_{\alpha}}{\partial x_{k}\partial x_{l}}}_{=0}\Bigg)_{2f\times2f}
        =D^{2}H^{(\sigma)}    
        \label{eq:D2L}
    \end{eqnarray}
    determines the searched index $r$ of the stationary point. 

    We restrict the Hessian $D^2L'$ to the subspace $\Sigma$ by taking the derivatives only in directions tangent to $\Sigma$ and omitting the derivatives in directions orthogonal to $\Sigma$, which correspond to gradients of the constraints $\nabla \Phi_{\alpha}(\vector{X})$ in the full space $\Omega$.
    The restricted Hessian
    \begin{equation}
        \label{eq:restrictedHessian}
        D^2L'|_{\Sigma} = \left(\begin{array}{cc}
            \left(D^{2}_{\vector{x}}L'\right)_{2f\times 2f} &(0)_{2f\times c} \\
             (0)_{c\times 2f} &(0)_{c\times c}
             \end{array}\right)
    \end{equation}
    has the same eigenvalue spectrum as $D^{2}_{\vector{x}}L'$ plus $c$ additional zero eigenvalues.
    The number of negative eigenvalues $r_\Sigma$ of $D^2L'|_{\Sigma}(\vector{X}'_{\mathrm{st}})$ equals $r$ of $D^{2}H^{(\sigma)}(\vector{x}_{\mathrm{st}})$, see~\eref{eq:D2L}, therefore it determines the index of the nondegenerate stationary point of $H^{(\sigma)}$.
    If the stationary point of $H^{(\sigma)}$ is degenerate, then the spectrum of the Hessian matrix has more than $c$ vanishing eigenvalues.
    
    In order to find an explicit form of the restricted Hessian~\eref{eq:restrictedHessian} we have to transform the coordinate basis at $\vector{X}_{\mathrm{st}}$ from the given Hamiltonian coordinates $\vector{X}=\left(X_{1},\dots,X_{2(f+c)}\right)$ to the adapted coordinates tangent and orthogonal to $\Sigma$.
    These coordinates do not necessarily have to be in the form~\eref{eq:xphiPhi}.
    Since $\nabla \Phi_\alpha(\vector{X}) \neq 0$ for every $\alpha$ and the vectors~$\nabla\Phi_{\alpha}(\vector{X})$ are linearly independent (otherwise the constraints would be ill-defined), the basis vectors for $c$ coordinates $(\xi_\alpha)_{\alpha=1}^{c}$ orthogonal to $\Sigma$ are obtained by orthonormalising the vectors $\{\nabla \Phi_\alpha(\vector{X})\}_{\alpha = 1}^c$, for example using the Gram-Schmidt orthogonalisation method.
    Then we complete the $d=2(f+c)$-dimensional orthonormal basis by an arbitrary set of mutually orthonormal vectors, all orthogonal to $\nabla \Phi_\alpha(\vector{X})$.
    These additional $2f+c$ basis vectors for coordinates $(\kappa_k)_{k=1}^{2f+c}$ are surely tangent to $\Sigma$, because the only vectors orthogonal to $\Sigma$ are the gradients of the constraints.
    A phase space point in the newly constructed basis has coordinates
    \begin{equation}
        \label{eq:NewCoordinates}
        \tilde{\vector{X}}=(\xi_{1},\dots,\xi_{c},\kappa_{1},\dots,\kappa_{2f+c}).
    \end{equation}
    If $M$ is the transformation matrix from the basis of $(X_{k})_{k=1}^{d}$ to the new basis, then
    \begin{equation}
        D^2L|_{\Sigma} = \left(M\cdot D^2L(\vector{X}_{\mathrm{st}}) \cdot M^{-1} \right)_{(c+1:d,c+1:d)}
    \end{equation}
    where $A_{(c+1:d,c+1:d)}$ represents a square submatrix of $A$ obtained by excluding the first $c$ rows and the first $c$ columns, and $\cdot$ denotes the matrix multiplication.
    Since the coordinates~\eref{eq:NewCoordinates} are not in the same basis as~\eref{eq:xphiPhi}, another similarity transformation is required to get the restricted Hessian exactly into the form~\eref{eq:restrictedHessian}; however, such a transformation will not change the spectrum of the Hessian eigenvalues, which determines of the index $r_{\Sigma}(\vector{X}_{\mathrm{st}})$.
    
\section{Properties of the Holstein-Primakoff mapping}\label{app:HPmapping}
This appendix reviews the theory of the classical HP mapping exposed in~\cite{Macek2019} and extends the theory of the HP mapping in the classical limit.

Starting from the boson operators $\B_{k},\Bc_{k}$, a quantum HP mapping is based on a new set of $f+1$ boson operators
\begin{equation}
    \label{eq:HPi}
    \bc{j}_{j}=\sqrt{\frac{\hat{N}}{\Bc_{j}\B_{j}}}\Bc_{j}, 
    \qquad \bc{j}_{k}=\B_{j}\sqrt{\frac{1}{\Bc_{j}\B_{j}}}\Bc_{k}\quad\textrm{for $k\neq j$},
\end{equation}            
which eliminates the boson of the type marked by the parenthesised superscript.
This boson encodes the conserving total number of excitations, $\bc{j}_{j}\b{j}_{j}=\op{N}$.
All $(f+1)^{2}$ elements $\Bc_{k}\B_{l}$ of the $\algebra{u}(f+1)$ algebra can be newly expressed by $f^{2}$ pairs of $\bc{j}_{k},\b{j}_{k},k\neq j$, only, which results in the mapping~\eref{eq:productOfTwoBBoperators}.
Therefore, the Hamiltonian $\H{j}$ expressed in the new set of boson operators does not explicitly contain any of $\bc{j}_{j},\b{j}_{j}$.

The classical limit is performed in a set of canonically conjugated positions $\q{j}_{k}$ and momenta $\p{j}_{k}$ introduced in the same way as in~\eref{eq:BosonicOpPQ}, leading to a classical Hamiltonian
\begin{equation}
    \label{eq:HPmapping}
    \lim_{N\rightarrow\infty}\frac{\H{j}}{N}\mapsto H(\vq{j},\vp{j}).
\end{equation}
   
An explicit form of the corresponding classical HP mapping in boson type~$j$, denoted $\mathcal{M}^{(j)}$, between original coordinates $(\vector{Q},\vector{P})$ and new coordinates $(\vector{q}^{(j)},\vector{p}^{(j)})$ is a canonical transformation that follows from~\eref{eq:HPi},
\begin{eqnarray}
    \q{j}_{j} = \frac{Q_{j} \sqrt{2}}{\norm{\vector{X}_{j}}}, \qquad  
    \p{j}_{j} = \frac{P_{j} \sqrt{2}}{\norm{\vector{X}_{j}}},\nonumber\\
    \q{j}_{k} = \frac{\vector{X}_{j} \cdot \vector{X}_{k}}{\norm{\vector{X}_{j}}}, \qquad 
    \p{j}_{k} = \frac{\vector{X}_{j} \cdot \symp \cdot \vector{X}_{k}}{\norm{\vector{X}_{j}}} \qquad \textrm{for $k\neq j$},
    \label{eq:qkpk}
\end{eqnarray}  
where $\vector{X}_{k}\equiv(Q_{k},P_{k})$ is a two-component vector built from a $k$-th pair of conjugated position and momentum, $\norm{\vector{X}_{k}}=\sqrt{\vector{X}_{k} \cdot \vector{X}_{k}}=\sqrt{Q_{k}^{2}+P_{k}^{2}}$ is the Euklidean norm of this vector and $\symp = \left(\begin{array}{cc} 0 & 1 \\ -1 & 0\end{array}\right)$ is the symplectic form. 
The presence of $\norm{\vector{X}_{j}}$ in the denominator makes the mapping $\mathcal{M}^{(j)}$ discontinuous at a region of a phase space with no $j$ bosons, $\norm{\vector{X}_{j}} = 0$.
The classical HP mapping~\eref{eq:qkpk} preserves the scalar product and the product with the symplectic form,
\begin{eqnarray}
\label{eq:conservedByHP}
    \vector{X}_{k} \cdot \vector{X}_{l} = \xv{j}_{k} \cdot \xv{j}_{l},\nonumber\\
    \vector{X}_{k} \cdot \mathfrak{W} \cdot \vector{X}_{l} = \xv{j}_{k} \cdot \mathfrak{W} \cdot \xv{j}_{l}\
    \quad\textrm{for $k,l \neq j$},
\end{eqnarray}
where $\xv{j}_{k}=(\q{j}_{k},\p{j}_{k})$.

A cyclic coordinate $\varphi^{(j)}$ conjugate to the integral of motion $R_{\Sigma}=\sqrt{2}$ can be introduced as $\q{j}_{j} = \sqrt{2}\cos\varphi^{(j)}$ and $\p{j}_{j} = \sqrt{2}\sin\varphi^{(j)}$, where $\varphi^{(j)}$ does not depend on the remaining coordinates.
Therefore, the integral of motion $R_{\Sigma}$ has been separated from the rest of the coordinates, and the nontrivial dynamics can be studied in the $2f$-dimensional reduced space $\s{j}$ spanned by coordinates $(\q{j}_{k},\p{j}_{k}),k\neq j$.

An inverse classical HP mapping $\left[\M{j}\right]^{-1}$ is obtained directly from~\eref{eq:qkpk}: 
\begin{eqnarray}
    Q_{j} = \q{j}_j\frac{\sqrt{2 - \sum_{k \neq j} \norm{\xv{j}_{k}}^{2}}}{\sqrt{2}},
    &P_{j} = \p{j}_j\frac{\sqrt{2 - \sum_{k \neq j} \norm{\xv{j}_{k}}^{2}}}{\sqrt{2}},\nonumber\\
     Q_{k} = \frac{\vector{x}_{j} \cdot \vector{x}_{k}}{\sqrt{2}},
    &P_{k} = \frac{\vector{x}_{j} \cdot \symp \cdot \vector{x}_{k}}{\sqrt{2}} \quad \textrm{for $k\neq j$}.
    \label{eq:inverseHP}
\end{eqnarray} 
  
Since two different mappings $\M{j},\M{j'},j\neq j'$ are canonical transformations, then the transformation $\mathcal{T}^{(jj')}$
between coordinates $\left(\vq{j},\vp{j}\right)$ and $\left(\vq{j'},\vp{j'}\right)$ is also canonical. 
It is explicitly given by formulas
\begin{eqnarray}
    \q{j'}_j= \q{j}_{j'}\frac{\sqrt{2 - \sum_{k \neq j} \norm{\xv{j}_{k}}^{2}}}{\norm{\xv{j}_{j'}}},\nonumber\\
    \p{j'}_j= -\p{j}_{j'}\frac{\sqrt{2 - \sum_{k \neq j} \norm{\xv{j}_{k}}^{2}}}{\norm{\xv{j}_{j'}}},\\
    \q{j'}_k=\frac{\xv{j}_{j'} \cdot \xv{j}_k}{\norm{\xv{j}_{j'}}},
    \quad\p{j'}_k=\frac{\xv{j}_{j'} \cdot \symp \cdot \xv{j}_{k}}{\norm{\xv{j}_{j'}}},\quad j\neq k\neq j'.\nonumber
\end{eqnarray}                       
Note that the term under the square root is the ratio of the numbers of particles $N_j/N_{j'}$, \cf Eq.~\eref{eq:nk}.
Therefore, $\mathcal{T}^{(jj')}$ is discontinuous at $N_{j'} = 0$ and the inverse transformation, $\left(\mathcal{T}^{(j'j)}\right)^{-1}\equiv\mathcal{T}^{(jj')}$, is not defined at $N_{j}=0$.

\end{appendices}
\vspace{1cm}

\input{refs.bbl}


\end{document}

%% file: refs.bbl
\providecommand{\newblock}{}